\documentclass[usenatbib,useAMS]{mn2e}

\usepackage{aas_macros}
\usepackage{xspace}
\usepackage{psfig}
\newcommand{\hh}{\ensuremath{\mathrm{H}_2}\xspace}
\newcommand{\um}{\ensuremath{\umu\mathrm{m}}\xspace}
\newcommand{\iso}{\textit{ISO}\xspace}
\newcommand{\thi}{2001ApJ...561.1074T}
\newcommand{\richter}{2002ApJ...572L.161R}

\title[A search for \hh around pre-main-sequence stars]{A search for \hh around pre-main-sequence stars}
\author[I.~Sheret, S.~K.~Ramsay~Howat and W.~R.~F.~Dent]
{I.~Sheret$^1$\thanks{E-mail: is@roe.ac.uk}, S.~K.~Ramsay~Howat$^2$ and W.~R.~F.~Dent$^2$\\
  $^1$Institute for Astronomy, University of Edinburgh, Royal
  Observatory, Blackford Hill, Edinburgh, EH9 3HJ \\        
  $^2$UK Astronomy Technology Centre, Royal Observatory, Blackford Hill, Edinburgh, EH9 3HJ}

\begin{document}

\date{Accepted 1988 December 15. Received 1988 December 14; in original form 1988 October 11}

\pagerange{\pageref{firstpage}--\pageref{lastpage}} \pubyear{2002}

\maketitle

\label{firstpage}

\begin{abstract}
  We present the results of a search for pure rotational molecular
  hydrogen emission from two pre-main-sequence stars, AB Aur and CQ
  Tau. Observations were made using MICHELLE, the mid-IR echelle
  spectrometer at the UK Infrared Telescope. We found some evidence
  for emission in the $J = 4 \rightarrow 2$ line in AB Aur, but no $J
  = 3 \rightarrow 1$ line from either star. We derive upper limits on
  line flux which are significantly smaller than previous line flux
  estimates based on \iso observations. This suggests that the
  emission detected by \iso is extended on scales of at least 6
  arcsec, and does not come from the disk as previously thought.
\end{abstract}

\begin{keywords}
circumstellar matter.
\end{keywords}

\section{Introduction}

Circumstellar disks play a vital part in both star and planet formation,
and if we are to understand these processes then it seems we must
understand the composition and physics of disks first.  The discovery
that disks can survive well into the main sequence lifetime of a star
(after planet formation ended in our own system) has opened up an
exciting window on the planet formation process
\citep{2001ARA&A..39..549Z}.  This means that we can directly probe the
environment of planet formation, even though planets themselves may be
undetectable.

The dominant constituent of the disks around pre-main-sequence stars
is \hh. However, observing emission from \hh which represents the bulk
of the warm ($\sim$200K) gas in the disks is not straightfoward.
Near-infrared \citep{2002ApJ...576L..73B}, UV emission
\citep{2000ApJS..129..399V} or absorption observations
\citep{2001ApJ...551L..97R} trace only the hottest gas or are
dependent on the line of sight through the cirumstellar disk.
Mid-infrared observations of the pure rotational emission lines of \hh
trace the bulk of the warm gas, but are difficult from ground-based
sites, due to telluric absorption and the high mid-infrared background
emission. Thus, many previous studies of the gas component of such
disks has relied on observations of tracer molecules, particularly CO
\citep{2000ApJ...529..391M,2000A&A...355..165D}. While these
observations have been successful in tracing the velocity structure in
disks, it seems that CO is a poor indicator of the total gas content.
This is because CO can be destroyed by UV photons or frozen onto
dust grains, leading to a severe underestimate of the total amount of
gas.  Most estimates of disk mass are based on sub-mm observations of
the dust continuum \citep{1990AJ.....99..924B,2001MNRAS.327..133S},
but these rely on assumptions about the composition and size
distribution of the dust grains, and also on the gas/dust mass ratio.
More importantly, we know that the dust and gas do not evolve in the
same way -- disks around main-sequence stars are composed mainly of
dust, while disks around pre-main-sequence stars are composed mainly
of gas. It is important to be able to measure gas properties
independently from the dust properties using direct observations of
\hh.

Recently, the Infrared Space Observatory (\iso) has been used to search
for \hh pure-rotational emission lines from pre-main-sequence stars,
and also young main-sequence stars which are known to have dust disks.
\citet{\thi} report detections of the \hh $J=3 \rightarrow 1$
transition ($\lambda = 17.035$ \um) and \hh $J=2 \rightarrow 0$
transition ($\lambda = 28.221$ \um). \hh masses derived from these
observations imply large masses of \hh in the disks around
main-sequence targets \citep{2001Natur.409...60T}, reversing the
previous belief that dust is the main component. However, as the \iso
beam is large (14 $\times$ 27 arcsec) compared with the disks and the
observations have moderate spectral resolution ($R \sim 2000$), these
results cannot distinguish circumstellar disk emission from that
of a foreground cloud or extended source.

Follow up observations of both the pure rotational emission lines and
\hh in absorption have provided strong evidence that the warm \hh is
extended in sources selected from the \iso studies. Using the the
echelle spectrometer, TEXES, \citet{\richter} obtained
observations of six sources, three of which have reported \iso \hh
detections \citep{\thi}: HD163296, AB Aur and GG Tau. They obtained
upper limits on the line fluxes from five sources, with a possible
2 sigma detection of the \hh $J=3 \rightarrow 1$ S(0) line in AB Aur
and conclude that the majority of the \hh emission detection by \iso
is extended on scales of 5 arcsec or more. The far-UV spectrum of Beta
Pictoris presented by \citet{2001Natur.412..706L} shows a complete
absence of \hh absorption lines, contrary to expectations if large
quantities of \hh are present within the edge-on dust disk.

We present the results of our attempts to confirm the \iso results
using MICHELLE, the mid-infrared echelle spectrometer on the
ground-based United Kingdom Infrared Telescope (UKIRT). We targeted
two pre-main-sequence stars, CQ Tau (distance 100 pc) and AB Aur
(distance 142 pc), both of which have good
\iso detections and one of which (AB Aur) was also observed by
\citet{\richter}. In both cases, the dust disk is expected to be
unresolved spatially by MICHELLE.

\begin{table*}
\begin{center}
\begin{tabular}{llllll}
\hline
Object   & Date        & Wavelength & Slit width  & Pixel scale on-sky  & Integration time \\
         &             & \um        & pixels & arcsec     & seconds     \\
\hline
AB Aur   & 29 Dec 2001 & 12.279   & 2  & 0.59$\times$0.24 & 1900   \\
         & 29 Dec 2001 & 17.035   & 2  & 0.50$\times$0.28 & 3400   \\
CQ Tau   & 30 Dec 2001 & 17.035   & 4  & 0.50$\times$0.28 & 5900   \\
\hline
\end{tabular}
\caption{Summary of observations. The pixel scale on the sky is subject
to anamorphic magnification due to the quasi-Littrow mode of operation
of the MICHELLE spectrometer. The long axis is along the slit.}
\end{center}
\end{table*}

\begin{figure}
\begin{center}
\psfig{figure=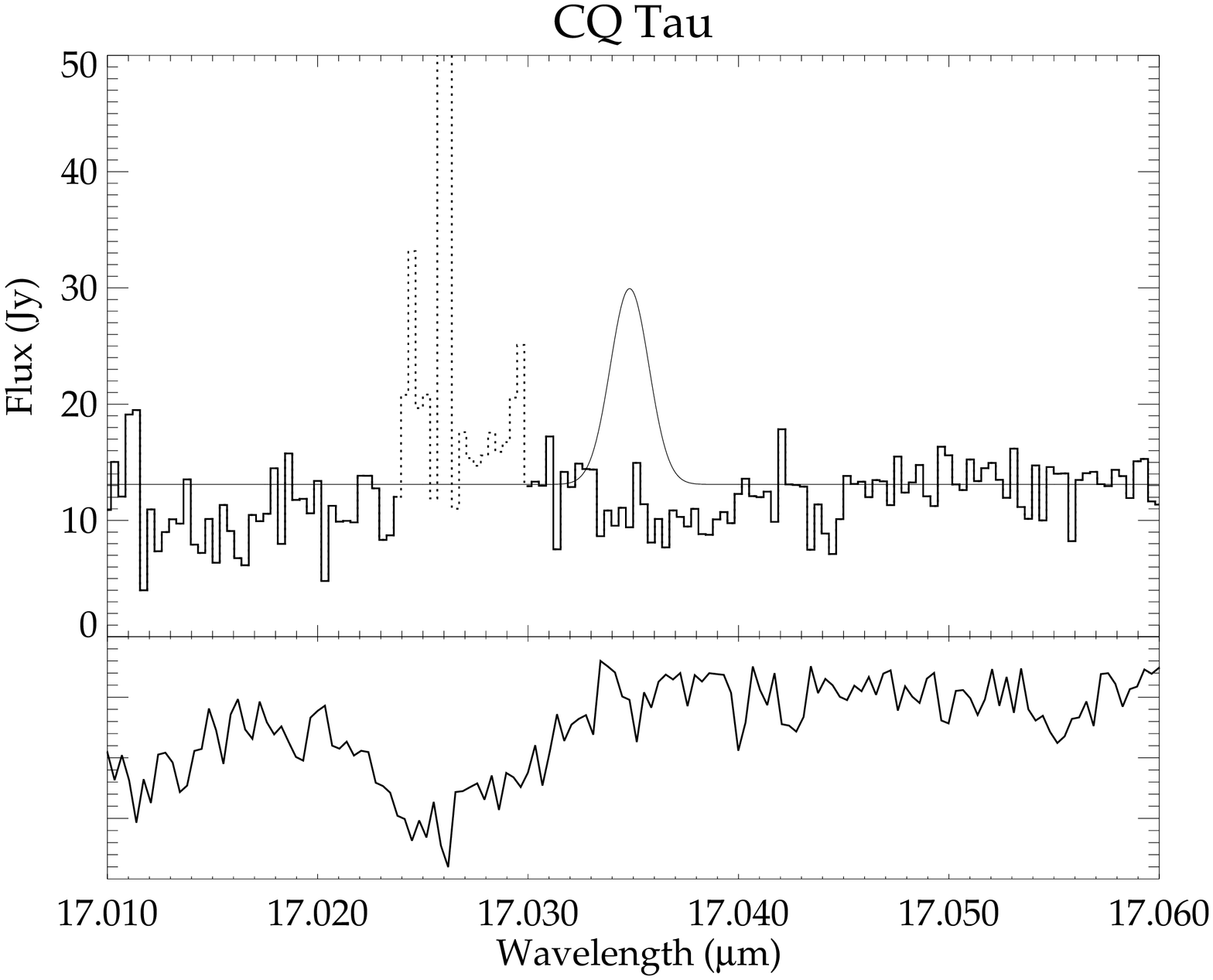,angle=90,width=\columnwidth}
\vspace{0.4cm}
\psfig{figure=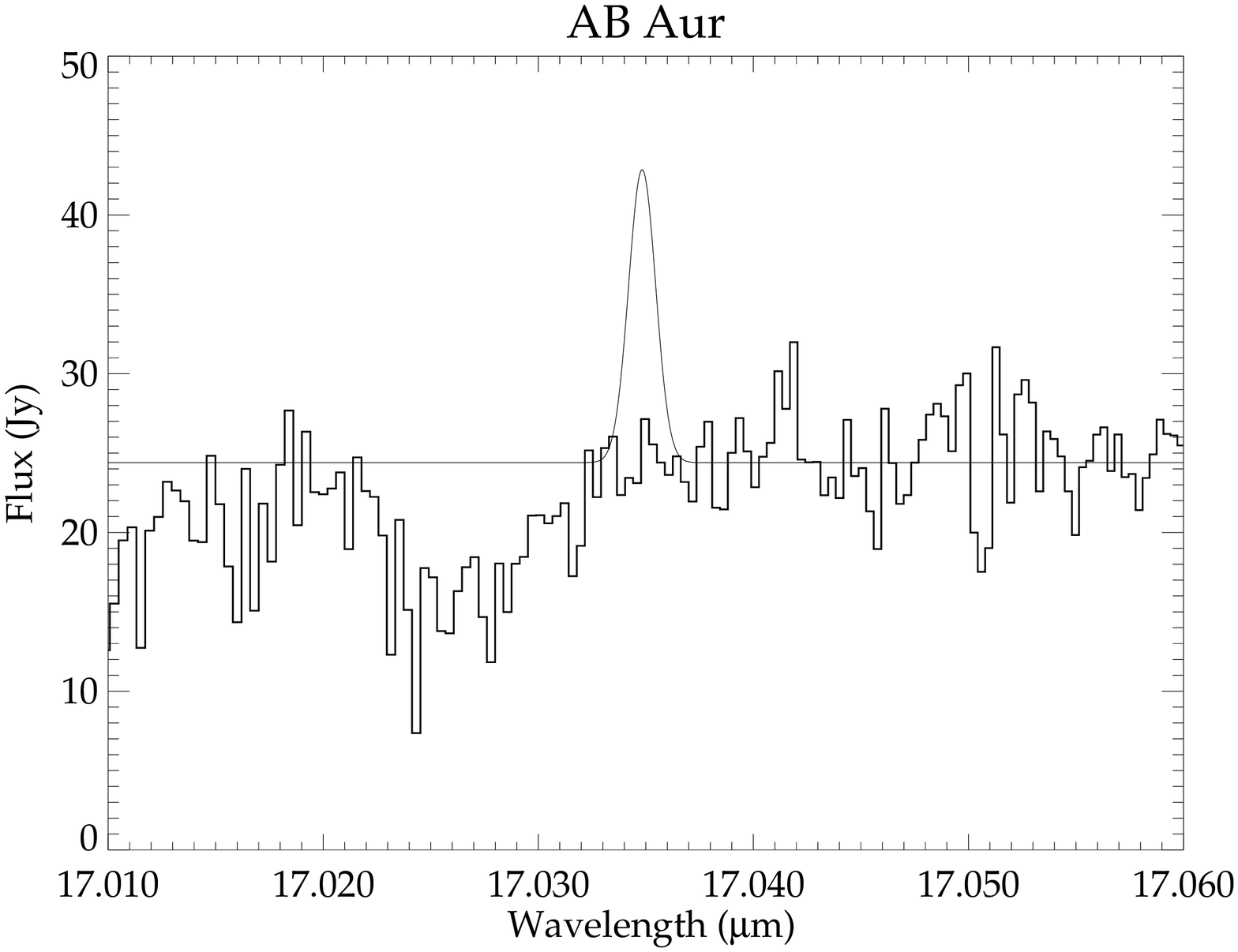,angle=90,width=\columnwidth}
\end{center}
\caption{Spectra of the \hh $J=3 \rightarrow 1$ transition for AB Aur
  and CQ Tau. For comparison, gaussians are shown with the same
  integrated flux as the \iso measurements reported in \citet{\thi},
  assuming an intrinsic line width of 20 km~s$^{-1}$. The CQ Tau
  spectrum has been masked out at regions of high noise (shown by the
  use of dotted lines); the lower panel shows the standard star
  observation, scaled from 0.0 to 1.0, to demonstrate that the masked
  regions correspond to atmospheric absorption lines.  For AB Aur, the
  standard star observation was saturated, which causes the
  variability in the continuum level.}
\end{figure}

\section{Observations and data reduction}

Mid-IR echelle spectroscopy of AB Aur and CQ Tau was carried out on
2001 December 29 and 30 using MICHELLE \citep{1997SPIE.2871.1197G}
on UKIRT. AB Aur and CQ Tau were observed in the \hh $J=3 \rightarrow
1$ transition ($\lambda = 17.035$ \um), providing direct comparison
with the \iso observations. AB Aur was also observed in the \hh $J=4
\rightarrow 2$ transition ($\lambda = 12.279$ \um). The \hh $J=2
\rightarrow 0$ transition ($\lambda = 28.221$ \um) was used in the
\iso observations, but is not readily accessible from the ground due
to the structure of the atmospheric transmission at 28 \um. Details of the
observations are shown in Table 1.

MICHELLE is a mid-IR imager and spectrometer which uses a
320$\times$240-pixel Si:As array and operates between 8 and 25
microns. For these observations the echelle grating was used, to
provide maximum sensitivity to the \hh lines which are expected to be
unresolved.  AB Aur was observed with a 2-pixel slit width; CQ Tau
with a 4-pixel slit width. The resolving power was $R=15\,200$ at 12
\um and $R=18\,600$ ($R=9\,300$) at 17 \um for a two (four) pixel wide
slit; as a result the \hh lines are well separated from the bright sky
lines. A single order from the grating is selected using blocking
filters. The pixel field of view varies with wavelength due to
anamorphic demagnification. The values are given in Table 1.

The point-like sources were all observed by nodding along the 70 arcsec
long slit, which was orientated north-south on the sky. The separation
of the two beams was 15 arcsec and the dwell-time in each beam was
40 seconds. 

Data reduction was done using software written in the \textsc{idl}
programing language. Pairs of frames from different nod positions were
subtracted, giving a set of difference frames each containing a
positive and a negative version of the source. In each frame, the
positive and negative versions were combined to give a single 2D
spectrum. Before the set of 2D spectra were combined into a final
image, the dataset was despiked using an iterative sigma cut. To do
this, all of the 2D spectra were compared, and datum were flagged as
bad if they were further than a specified number of standard
deviations from the mean of their pixel.  This sigma cut was done
three times with threshold settings of 6, 4 and 3 sigma.

\begin{table*}
\begin{center}
\begin{tabular}{lllll}
\hline
Object    & Wavelength & \multicolumn{3}{c|}{Line flux}\\
          & \um        & \multicolumn{3}{c|}{$\times 10^{-14} $ ergs s$^{-1}$ cm$^{-2}$}\\
\cline{3-5}
          &            & Wide extraction & Optimal extraction & \iso Result\\
\hline                                                         
AB Aur    & 12.297     & 5.4 $\pm$ 3.8   & 4.8 $\pm$ 2.1      &    \\
          & 17.035     & $<$ 14.6        & $<$ 8.9            & 30 $\pm$ 9 \\
CQ Tau    & 17.035     & $<$ 13.0        & $<$ 6.7            & 40 $\pm$ 12\\
\hline
\end{tabular}
\end{center}
\caption{Summary of results. All upper limits are 3 sigma. \iso
  results are taken from \citet{\thi}. }
\end{table*}

Once a final 2D spectrum had been produced, 1D spectra were extracted
in two ways: a wide extraction which summed all the emission in a 6.5
arcsec region centered on the source, and an optimal extraction of the
continuum. The optimal extraction method takes a weighted mean
of the pixels around the source, giving more weight where the continuum is
brighter. This method gives the lowest possible noise on the continuum
level, but if the \hh emission region is more extended than the
continuum emission then it will underestimate the \hh line flux. For
both extraction methods a sky subtraction was done first, estimating
the residual sky signal in each column using pixels between 3.5 and 7
arcsec from the source.

Standard stars were observed and reduced in the the same way as the
science targets. The final spectrum for each object was divided
through by a standard spectrum to remove wavelength dependant
variations in atmospheric and instrumental transmission. The standard
stars were observed using the same rows on the detector as the
sources, and division by the standard effectively flat-fields the
detector. For the 17 \um observations of AB Aur the standard star
observation was saturated, so we were unable to make this correction.
This causes variation in the continuum level, but would not prevent us
from detecting a narrow emission line.

Wavelength calibration was obtained by measuring the positions of sky
lines, and comparing the positions to a model sky spectrum. We
estimate a 1 sigma error on the wavelength calibration is about one
pixel ($\sim 3 \times 10^{-4}$ \um), and is largely due to the
difficulty in centroiding the broad sky lines. The spectra were
shifted to the object's rest frame, accounting for the radial velocity
of the earth and object's heliocentric radial velocity (+8.9 km
s$^{-1}$ for AB Aur and +22.6 km~s$^{-1}$ for CQ Tau
\citep{1999yCat.3213....0B}). Flux calibration was based on the
observed continuum level and published continuum flux measurements
from \citet{\thi} for both the 17 \um spectra and \citet{\richter} for
the AB Aur 12 \um spectrum.

\begin{figure}
\begin{center}
\psfig{figure=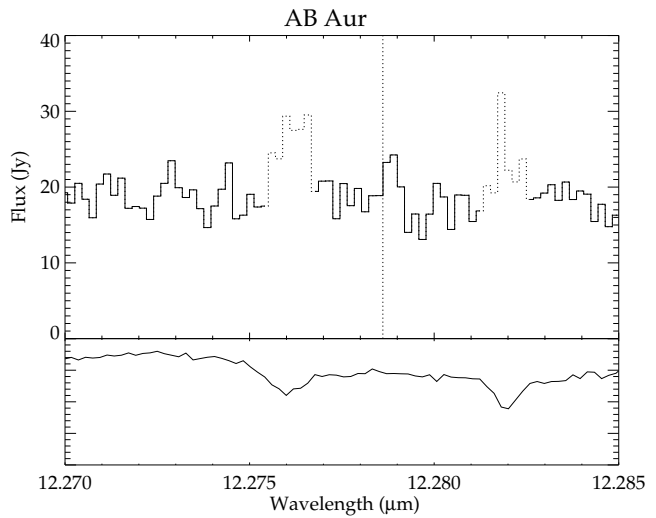,angle=90,width=\columnwidth}
\end{center}
\caption{Spectra of the \hh $J=4 \rightarrow 2$ transition for AB
  Aur. The vertical dotted line shows to the expected location of the
  transition. Noisy regions have been masked out and the standard star
  is shown in the lower panel, as in Figure 1.}
\end{figure}

\section{Results}

Figures 1 and 2 show the final spectra produced using an optimal
extraction.  At the location of sky lines the noise increases
dramatically, so these regions have been masked out (i.e. they were
not used when estimating the continuum level and the overall noise).
On the 17 \um figures a gaussian with the total line flux
measured by \citet{\thi} is shown, to allow comparison between the
\iso results and our spectra. As the \iso observations do not resolve
the line, we have assumed an intrinsic line width of 20 km~s$^{-1}$
(see Section \ref{SEC:DISCUSSION} for discussion of this value), and
an instrumental width of 16 km~s$^{-1}$ for AB Aur and 32 km~s$^{-1}$
for CQ Tau, giving a total width of 26 km~s$^{-1}$ and 38 km~s$^{-1}$
respectivley. The MICHELLE sensitvities should have produced a clear
detection of the line, if all of the flux in the $14 \times 27$ arcsec
\iso beam were contained within our extracted beams, as expected for
emission from an unresolved disk.

To estimate the line flux, all of the flux within a 40~km~s$^{-1}$
wide bin centered on the systematic velocity was summed, after
subtracting the continuum level. Fluxes and upper limits for all
extraction methods are shown in Table 2.  There is no evidence of \hh
emission in either of the 17 \um spectra.  The 12 \um optimally
extracted spectrum gives flux of $4.8 \pm 2.1$ ergs s$^{-1}$
cm$^{-2}$, but as this is only a 2.3 sigma result should not be
regarded as a definite detection. However, this result is consistent
with the tentative detection by \citet{\richter}, who measure $(2.0
\pm 1.0) \times 10^{-14}$ ergs s$^{-1}$ cm$^{-2}$ with a width of 10.5
km~s$^{-1}$.

\section{Discussion \label{SEC:DISCUSSION}}

Despite the enhanced sensitivity to narrow emission lines from a
point-like source, these new spectra of AB Aur and CQ Tau show no
convincing detections of either the 0-0 S(1) or 0-0 S(2) lines. In
order for the 0-0 S(1) \hh emission to be detectable by \iso but not
by MICHELLE, the emission would have to be either spectrally broad or
spatially extended. We consider these two explanations in turn.

The spectral resolving power of the \iso observations is ${\rm R}~(
=\frac{\lambda}{\delta\lambda})=2400$, and the \hh lines \citet{\thi}
observed are unresolved. In calculating our upper limits, we assume
that the lines had an intrinsic width of 20 km~s$^{-1}$. The dominant line
broadening mechanisms for gas in a disk are thermal broadening, and
bulk motion due to the disk's rotation. Assuming the gas has a
temperature of 200 K (as estimated by \citet{\thi}), the thermal
broadening would only be 1.5 km~s$^{-1}$. The amount of rotational
broadening depends on the distance from the star, and the
inclination of the disk. If we assume a distance of 2AU (a reasonable
choice, given the gas temperature of 200 K), this gives a width of 15
km~s$^{-1}$, assuming an inclination of $\sim45^\circ$.  Therefore, 20
km~s$^{-1}$ is a conservative estimate of the line width. In fact, the
line FWHM would have to be over 100 km~s$^{-1}$ to make the \iso and
MICHELLE measurements consistent.

The possibility of extended emission, due to excitation of a remnant
cloud of molecular gas by UV photons or by shocks, is plausible.
Assuming that the \iso measured flux is uniformly distributed over the
$14\times27$ arcsec aperture, the fluxes are consistent with
UV-excited emission strengths predicted by \citet{1992ApJ...399..563B}
for moderate UV field strengths ($10^{3}-10^{4}$ times that of the
interstellar medium) and moderate densities ($n_{\rm
  H_{2}}\sim10^{3}-10^{4}{\rm cm}^{-3}$). Over the MICHELLE beam used
to obtain the spectra shown in Figure 1, the line flux would be an order
of magnitude fainter than the sensitivity limit obtained. Moreover, the
sky subtraction region (3.5-7 arcsec radius) means that our
observations are insensitive to emission at radius $>$3.5 arcsec. The
hypothesis of extended emission is strengthened for AB Aur since a map
of $^{12}$CO 3-2 shows an extended molecular cloud around the central
source, with a size of at least 30 arcsec \citep{\thi}.  For CQ Tau,
no $^{12}$CO 3-2 line was detected, but this could be
reconciled with extended \hh emission if the CO has been destroyed by
photodissociation \citep{1995MNRAS.277L..25D}.

In interpreting our results, an important question concerns the
optical depth of the dust disk. If the dust is optically thick at the
wavelengths we observed and the gas and dust are at the same
temperature, then the gas would be equally likely to absorb or emit a
photon, so no emission line would be produced. Line emission will only
occur if there are gaps in the dust disk, or if there is a hot
atmosphere above the disk. Though these situations are quite
plausible, only a small fraction of the gas in the disk would be
observed in this case, and any calculation of the gas mass
would underestimate the true mass. It is therefore essential that we know
the optical depth of the disk in order to correctly interpret our
results.

The optical depth of the disks can be estimated from radiative
transfer models based on the SED of the disks.
\citet{2001A&A...371..186N} made models of the disk SEDs for both AB
Aur and CQ Tau, and we can use their fitted parameters to determine
whether the disks are optically thick at 12 and 17 \um. Based on their
estimates of disk mass and disk radius, and using the usual assumption
that the dust opacity at 1.3 mm is $\kappa_{1.3 \mathrm{mm}} = 0.01$
cm$^2$ g$^{-1}$, the average optical depth of the disks at 1.3 mm is
0.03 for AB Aur and 0.24 for CQ Tau.  Then, using a power law dust
opacity index ($\kappa \propto \lambda^{-\beta}$, where $\beta$ was
estimated to be 2.0 for AB Aur and 1.0 for CQ Tau in
\citealt{2001A&A...371..186N}), the optical depth can be estimated at
shorter wavelengths. This indicates that the disks become optically
thick at wavelengths shorter than 200 \um for AB Aur and 300 \um for
CQ Tau.  Both disks must therefore have a high optical depth at 12 and
17 \um.

This result has two important implications. Firstly, this means that
we cannot use our upper limits on line flux to calculate upper limits
on the mass of \hh in the disk, as most of the gas in the disk is
hidden by the optically thick dust. The \hh line at 12 \um (if
confirmed) must arrise from a hot atmosphere above the disk, or from
gaps in the dust disk. Secondly, if the the lines detected by \iso do
arise in the disk, then only a small fraction of the gas was observed
and the total gas mass must be much higher than was estimated.

In conclusion, we have searched for pure rotational \hh emission lines
from the circumstellar disks of AB Aur and CQ Tau. We tentatively
detect emission from AB Aur, but find no evidence for emission from CQ
Tau. The likely reason for this non-detection is that the disks are
optically thick in the mid-IR. Our upper limits on line flux are
significantly smaller than previous line flux estimates based on \iso
observations, which suggests that the emission detected by \iso is
extended on scales of at least 6 arcsec (or $>$ 60AU). 

Mapping the emission over the area covered of the \iso beam is one
approach to confirming these conclusions. Detecting the faint extended
emission will require a considerable gain in sensitivity over these
observations, such as is anticipated using MICHELLE on the Gemini
North 8-m telescope.

\vspace{7mm}

The United Kingdom Infrared Telescope is operated by the Joint
Astronomy Centre on behalf of the UK Particle Physics and Astronomy
Research Council. We wish to thank the staff at UKIRT for enabling
these observations. We also thank Mark Wyatt and Alistair Glasse for
helpful discussions of this work.

\bibliographystyle{mn}
\bibliography{michelle_gas}

\bsp

\label{lastpage}

\end{document}